\documentclass[a4paper,12pt]{article}

\usepackage{graphicx,epsfig,natbib}

\usepackage{amssymb}

\usepackage{epsfig}

\usepackage{subfig}

\usepackage{amssymb}

\usepackage{float}
\restylefloat{table}

\begin{document}


\begin{center}
\LARGE ``SCINDA-Iono'' toolbox for MATLAB: \\
analysis of ionosphere scintillations
\end{center}

\

\centerline{T.~Barlyaeva, T.~Barata, A. Morozova}

\begin{center}
 \it\small
 CITEUC, University of Coimbra, Portugal \\
SWAIR project
\end{center}

\vspace{0.7cm}

\begin{abstract}
Here we present a ``SCINDA-Iono'' toolbox for the MATLAB.
This is a software to analyze ionosphere scintillation indices provided by a SCINDA GNSS receiver.
The toolbox is developed in the MATLAB R2018b.
This software allows to preprocess the original data and analyze ionosphere scintillations on the 1-minute and 1-hour time scales both for
averaged over all available satellites values and separately for each receiver-satellite pair.
\end{abstract}

\begin{center}
\end{center}


\newpage

\section*{Current code version}
\label{}


\begin{table}[H]
\begin{tabular}{|l|p{6.5cm}|p{6.5cm}|}
\hline
\textbf{Nr.} & \textbf{Code metadata description} & \\
\hline
C1 & Current code version & v1.0.1 \\
\hline
C2 & Permanent link to code/repository & $https://fr.mathworks.com/$ \\
   & used for this code version        & $matlabcentral/fileexchange/$ \\
   &                                   & $71784-scinda-iono\_toolbox$ \\
\hline
C3 & ``SCINDA-Iono'' toolbox compute & $https://fr.mathworks.com/$ \\
   & capsule                         & $matlabcentral/fileexchange/$ \\
   &                                 & $71784-scinda-iono\_toolbox$ \\
\hline
C4 & Legal Code License   & MathWorks File Exchange: T.~Barlyaeva personal license \\
\hline
C5 & Code versioning system used & none \\
\hline
C6 & Software code languages, tools, and services used & MATLAB \\
\hline
C7 & Compilation requirements, operating environments \& dependencies & MATLAB R2018b \\
\hline
C8 & If available Link to developer & $https://fr.mathworks.com/$ \\
   & documentation/manual           & $matlabcentral/fileexchange/$ \\
   &                                & $71784-scinda-iono\_toolbox$ \\
\hline
C9 & Support email for questions & TVBarlyaeva@gmail.com \\
\hline
\end{tabular}
\caption{Code metadata} 
\label{}
\end{table}


\section{Motivation and significance}
\label{}

Nowadays, such technologies as GNSS navigation, radio- and telecommunications are widely used in various fields of human life
and, moreover, make an important part of our life. That is why the integrity of these systems is very important:
not only for the everyday comfort but, and that is more important, for the safety of life.

The quality of the GNSS signal depends on the conditions of the medium to pass, in particular on the conditions of the ionosphere.
There are a lot of natural factors that can affect quality of the used technical equipment's work. So, an investigation of the ionosphere,
its scintillations and conditions under various affecting factors is very important for a development, for instance, space weather alerts' systems for GNSS air navigation.
Knowledge of the factors influencing ionosphere can help to forecast a so called loss-of-lock of satellites and, thus,
to estimate possible position errors for GNSS navigation systems.
That is why the studies of ionosphere are highly needed nowadays: both from the scientific and practical points of view.

Measurements of ionosphere conditions provided by GNSS receivers are a very good source of the observational ionosphere data.
The analysis of these data can make an important contribution to the ionosphere science, to improve the integrity of such systems as GNSS
and, as consequence, to serve for the safety of life.

As a rule, each receiver model uses its special software and the output data can differ by their formats.
The most often are either ASCII format RINEX or its binary equivalent, BINEX.
But in all cases, one needs a software allowing to make an analysis of ionosphere data provided by a GNSS receiver easier and faster.
Here we present a MATLAB toolbox created to work with the SCINDA GNSS receiver ionosphere scintillation data in the ASCII format.
As we know, for now there is no tool allowing to preprocess/process/analyze such data in MATLAB, so we hope that our toolbox can be useful and demanded in the ionosphere science community.

The ionosphere scintillation measurements by the SCINDA receiver are described in detail in \cite{Carrano2007}--\cite{Carrano2009c}
with particular attention to estimations of the Total Electron Content (TEC) calculations \cite{Carrano2009a,Carrano2009b}.

The ``SCINDA-Iono'' toolbox presented here was developed to make an analysis of the SCINDA ionosphere data more comfortable for a user.
The only requirement for this toolbox is the MATLAB R2018b or higher.
Since MATLAB is a tool widely used in the scientific community, this toolbox can be useful and easily adapted for scientific studies of the ionosphere.

\section{Software description}
\label{Descr}

The presented toolbox is designed to work with the ASCII data files generated
by the SCINDA GNSS receiver. All the examples of the toolbox application to
the data analysis presented in this paper are based on the data acquired by a GNSS receiver
installed at the Lisbon airport (Portugal) at the end of 2014. The
installed equipment is a NovAtel EURO4 with a JAVAD Choke-Ring antenna.
The installed Firmware (SCINDA) is specific for scintillation detection.

The original SCINDA scintillation files have an extension {\it .scn} and are provided in the zipped form.
The original zipped files are of about $10kb$, the unzipped files are of about $30-45kb$ size.
Each {\it .scn} file consists of 1-minute resolution data, and contains data
for the pairs of the receiver with the all available for that moment satellites.
Thus, each row of the data corresponds to the various satellites whose `PRN' numbers are indicated in last column of the file.

\vspace{0.5cm}
The parameters listed in the files are (column by column from left to right):

\begin{itemize}
\item 2-digit year ($YY$),
\item month ($MM$),
\item day ($DD$),
\item seconds since midnight ($UTSEC$),
\item azimuth in degrees ($AZ$),
\item elevation in degrees ($EL$),
\item scintillation intensity index on L1 ($L1S4$),
\item scintillation intensity index on L2 ($L2S4$),
\item differential pseudorange ($TEC_P$),
\item differential carrier phase ($TEC_\Phi$),
\item relative (uncalibrated) TEC in TECU ($TEC_R$),
\item time since last cycle slip ($N$),
\item pseudorandom noise satellite identifier ($PRN$).
\end{itemize}

\vspace{0.5cm}

The ``SCINDA-Iono'' toolbox allows:
\begin{itemize}
\item to unzip scintillation data (.scn) provided by SCINDA GNSS receiver,
\item to extract the data obtained using various receiver-satellite pairs,
\item to preprocess these data and
\item to analyze the 1-min and 1-hour resolution data for the averages over all satellites and for each receiver-satellite pair separately.
\end{itemize}

\vspace{0.5cm}
The toolbox does an analysis of the data in the time interval chosen inside one calendar month.

\subsection{Software Architecture}
\label{Arch}

\begin{figure}
\noindent
\centering
\includegraphics[width=0.91\textwidth]{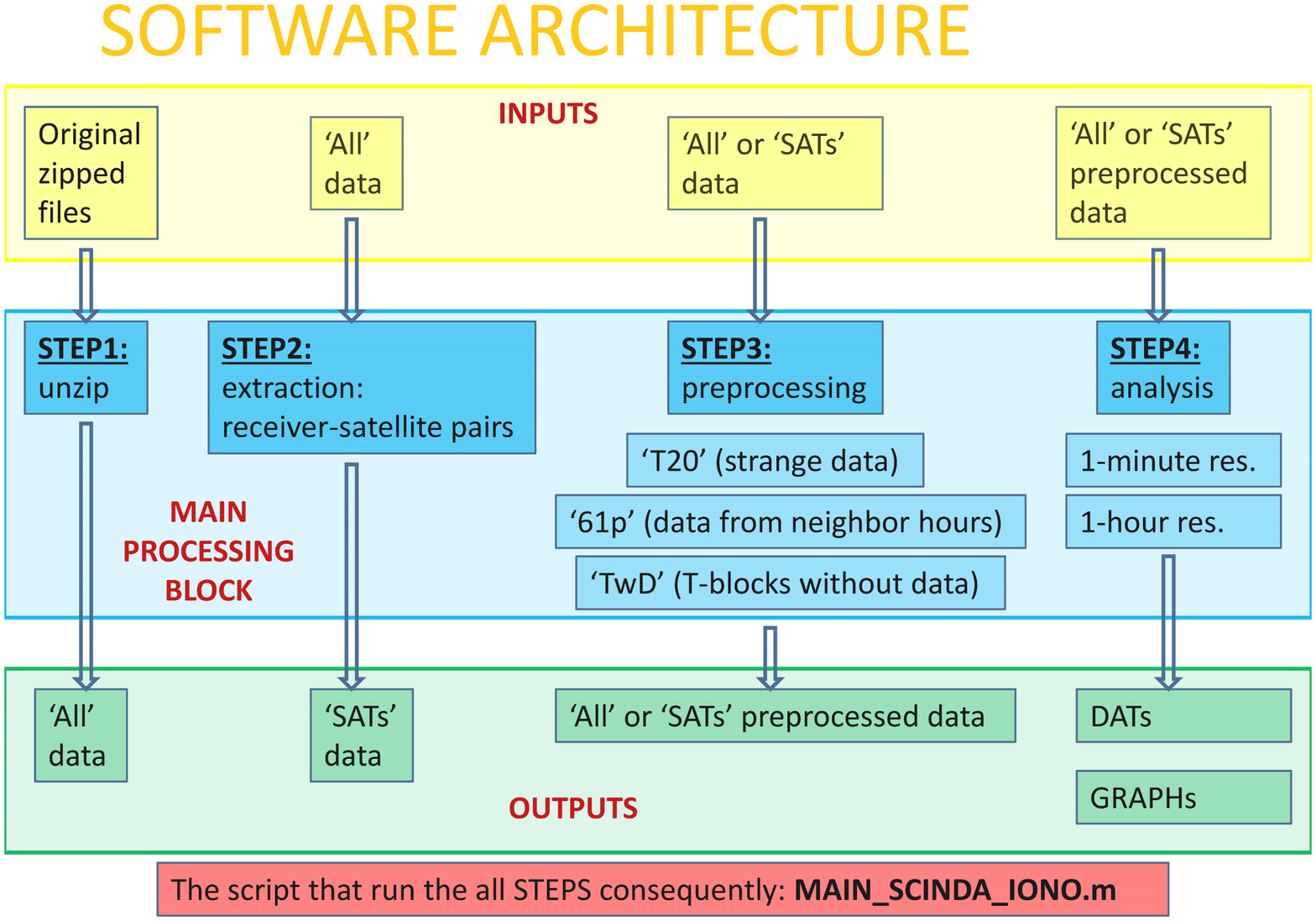}
\caption{Software architecture.}
\label{fig:Arch}
\end{figure}

The architecture of the ``SCINDA-Iono'' toolbox is schematically presented in Fig.~\ref{fig:Arch}.
It can be formally divided into three principal blocks:

\begin{itemize}
\item an adaptation of the receiver provided scintillation data for the further analysis;
\item their preprocessing;
\item analysis.
\end{itemize}

The preprocessing includes the corrections of the data for the receiver's miscalculations and the internal clock failures.
There are three types of the preprocessing procedures (see Figure~\ref{fig:Arch}).

\subsection{Software Functionalities}
\label{Funct}

The ``SCINDA-Iono'' toolbox allows to do the following operations:

\begin{itemize}
\item {\it To unzip the scintillation data.} Since the original data are provided in the archived format, to be processed they should be unzipped.
\item {\it To create the data files of the each receiver-satellite pair}, that have the format similar to the one of the original scintillation data files.
\item {\it To delete epochs with erroneous values (so-called `T20' preprocessing).}
Some epochs starts as normally from the letter `T', but instead of the two-digits year the `-20' value appears.
Even if content of such epochs seems to be typical we prefer to delete such epochs with erroneous headers.
\item {\it To put all data from an hour to the file containing the data on this hour (so-called `61p' preprocessing).}
Some hourly files of 1 minute data contain one measurement from the `neighbour' hour, thus the 61th measurement appears.
Such `extra' measurements are moved to the corresponding hour file.
\item {\it To delete the epochs without data, that contains only titles (so-called `TwD' preprocessing).}
Some blocks are empty: they don't contain any data except the header.
\item {\it To analyze the 1-minute resolution data.} As a result, the graphs of all preprocessed data are plotted.
\item {\it To analyze 1-hour resolution data.} Here the 1-hour means are accompanied by the standard deviation (Std)
and number of successful observations (Numb. of Points).

It should be noted that 1) the data obtained on the each step listed above are saved into indicated folders and
2) each of the steps listed above can be included or excluded by a user upon decision and specification of a user's task.
In latter case, one has to pay attention to the names of the files to be used for the further steps.
\end{itemize}

\section{Illustrative Examples}
\label{Examples}

Here we present an example of the ``SCINDA-Iono'' toolbox work
using the data from the SCINDA GNSS receiver installed in Lisbon airport (Portugal).
We preprocess and analyze the data for December, 2014. This interval was chosen at random,
without any special restrictions on ionosphere conditions.

\begin{figure}
\noindent
\centering
\includegraphics[width=0.91\textwidth]{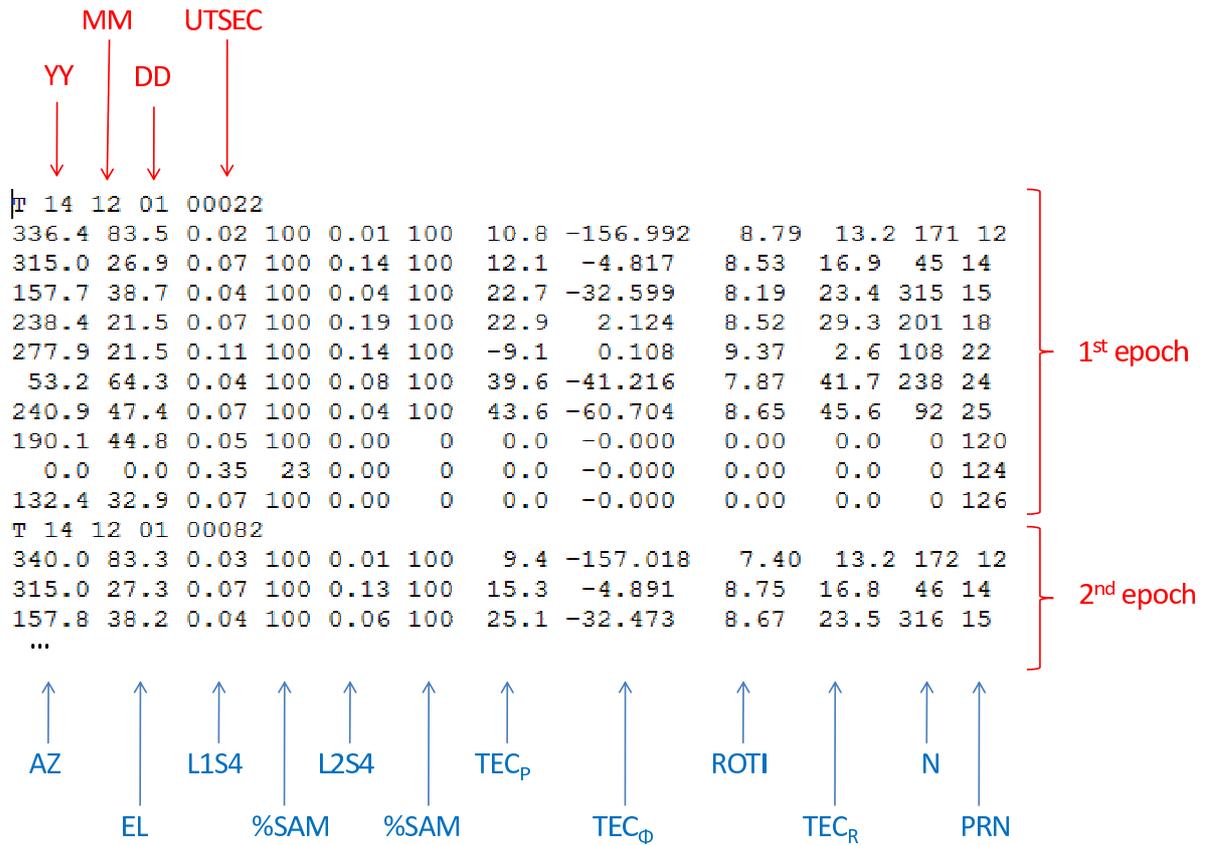}
\caption{Example of an original scintillation data file.}
\label{fig:Orig}
\end{figure}

The original scintillation data are of the format presented in Fig.\ref{fig:Orig}.
This data file is an input file for the ``SCINDA-Iono'' toolbox. This file contains both satellite characteristics and ionospheric parameters.
Each file consists of values registered by various satellites, and each line of the file contains data
obtained by one receiver-satellite pair. The toolbox is developed for the original data of 1-minute resolution.

\begin{figure}
\noindent
\centering
\includegraphics[width=0.91\textwidth]{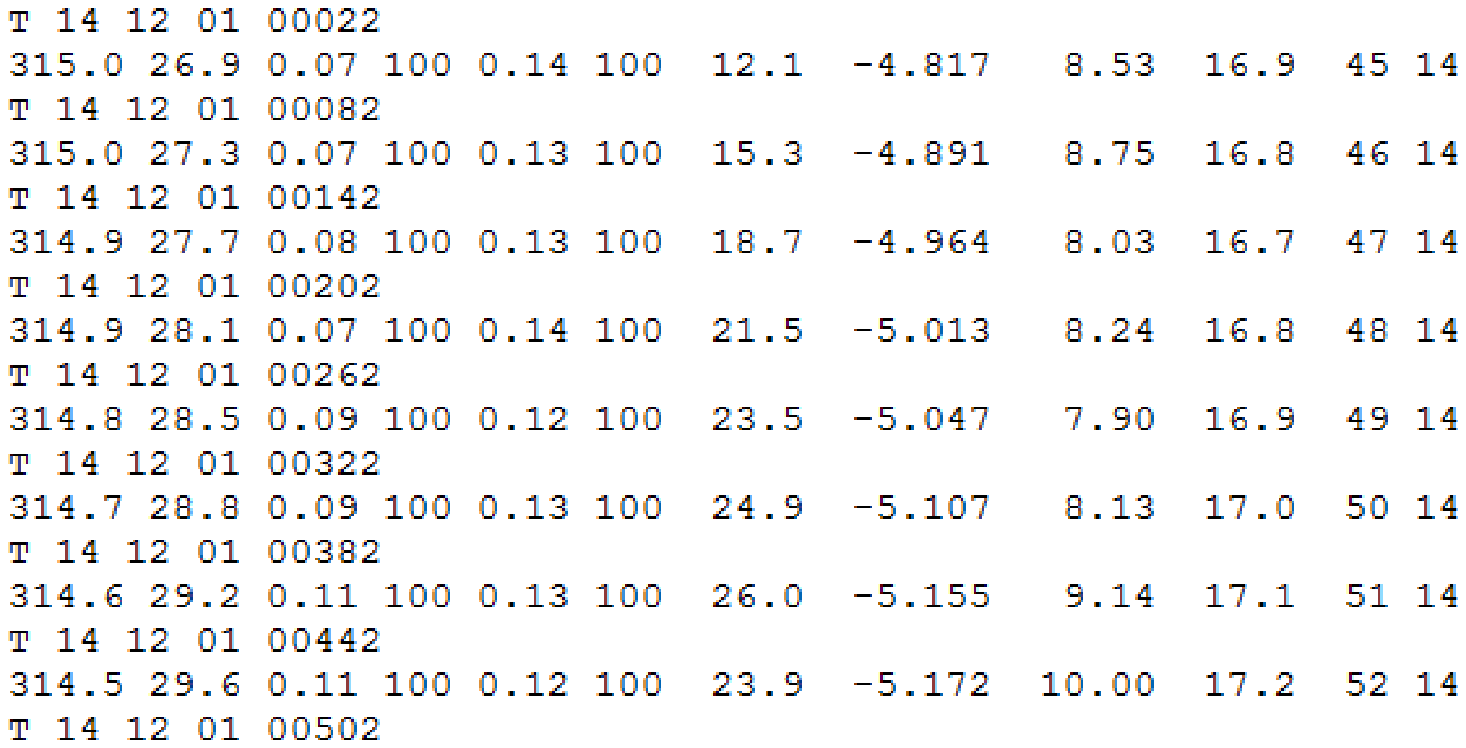}
\caption{Example of a scintillation data file for one receiver-satellite pair.}
\label{fig:SATs}
\end{figure}

\begin{figure}
\noindent
\centering
\includegraphics[width=0.73\textwidth]{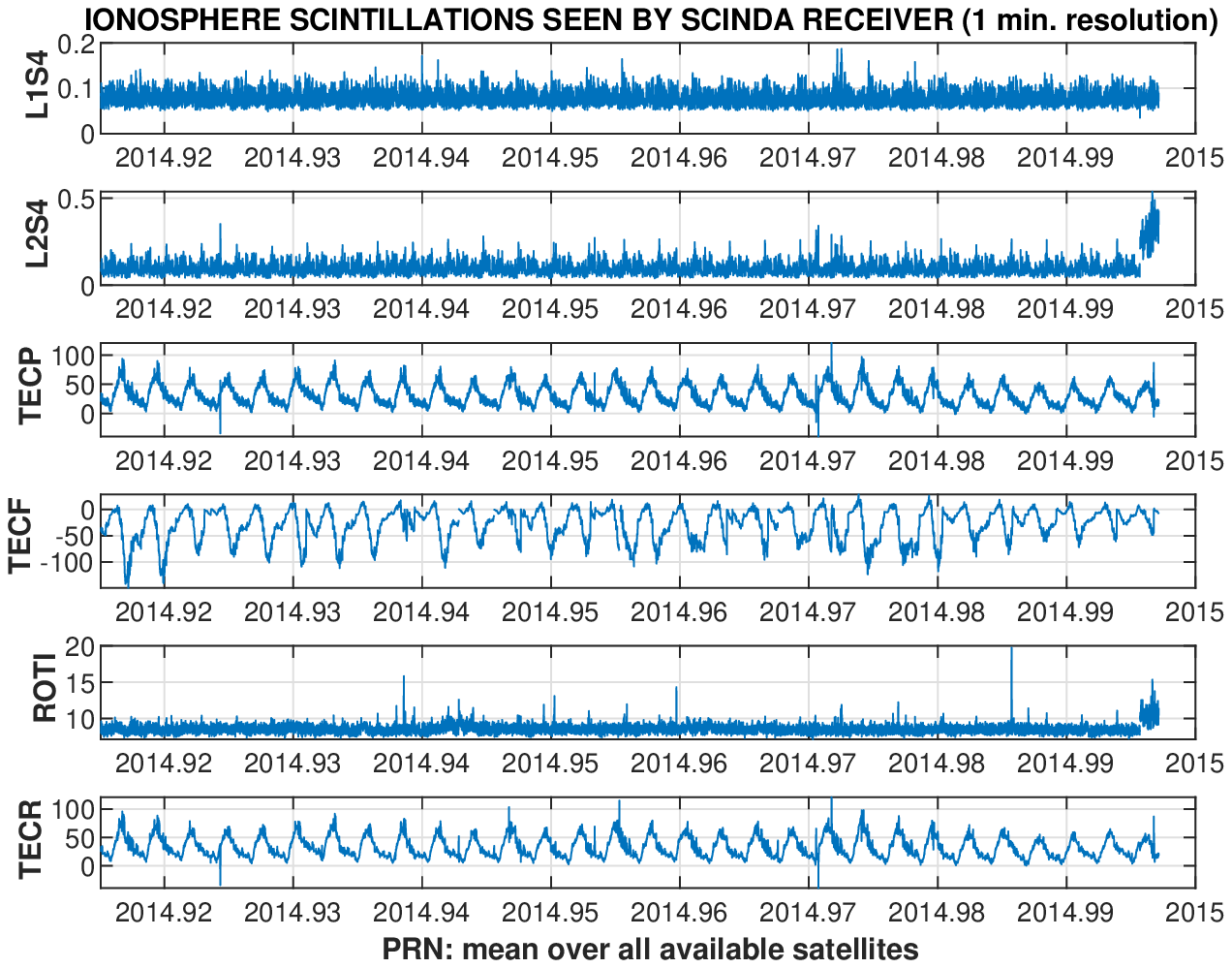}
\vspace{0.5cm}
\caption{Example of 1-minute resolution plots for an average over all available satellites.}
\label{fig:Mean1m}
\end{figure}

As was already mentioned above the software can do an extraction of the data for each of such receiver-satellite pairs.
Also, the data for the averages over all available for the moment satellites can be calculated.
An example of the resulting data file for the latter (``mean'') case is presented in Fig.~\ref{fig:SATs}.
The corresponding 1-minute resolution plots for an average over all available satellites are presented in Fig.~\ref{fig:Mean1m}.

\begin{figure}
\noindent
\centering
\includegraphics[width=0.73\textwidth]{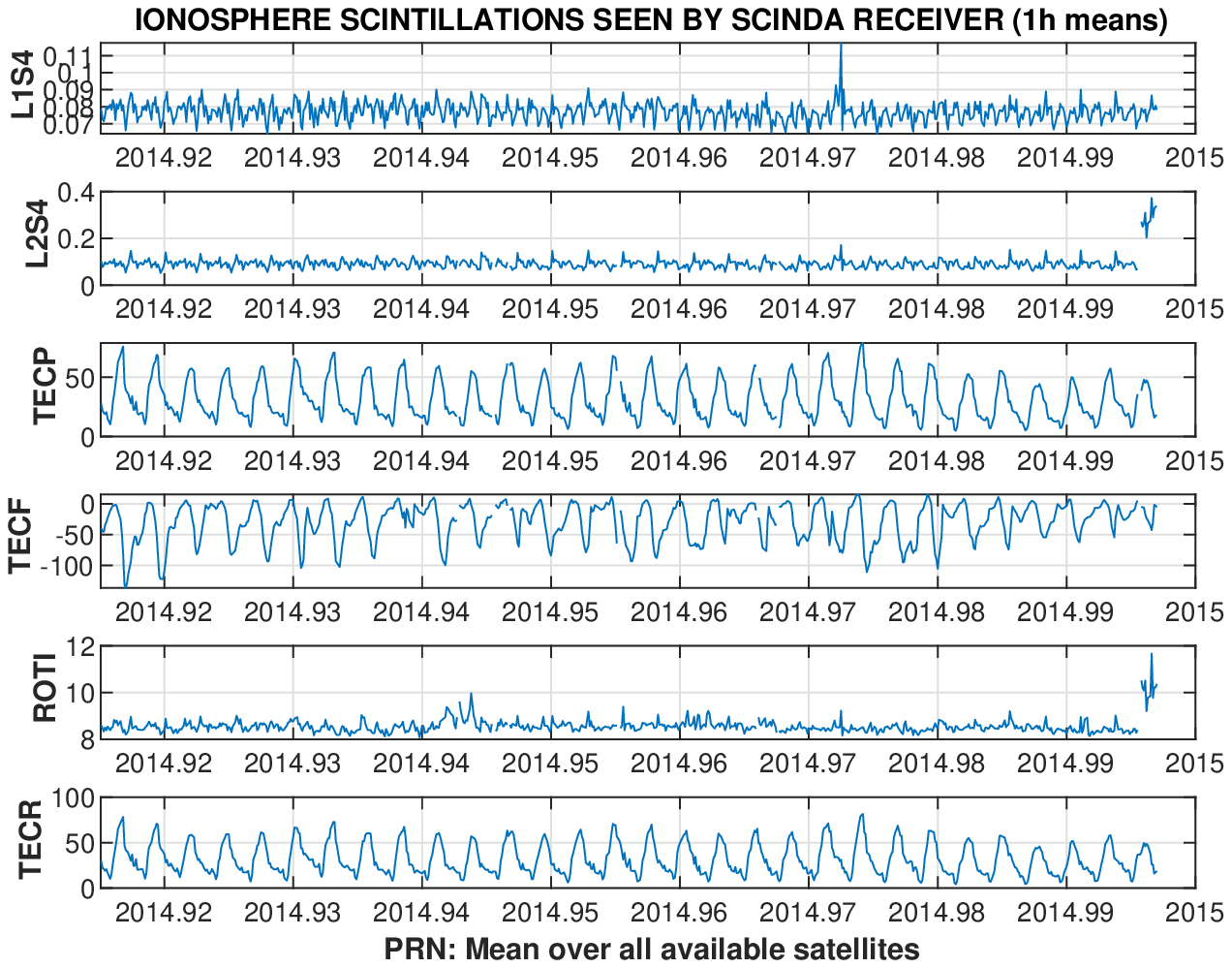}
\vspace{0.5cm}
\caption{Example of 1-hour resolution plots for an average over all available satellites.}
\label{fig:Mean1h}
\end{figure}

\begin{figure}
\noindent
\centering
\includegraphics[width=0.73\textwidth]{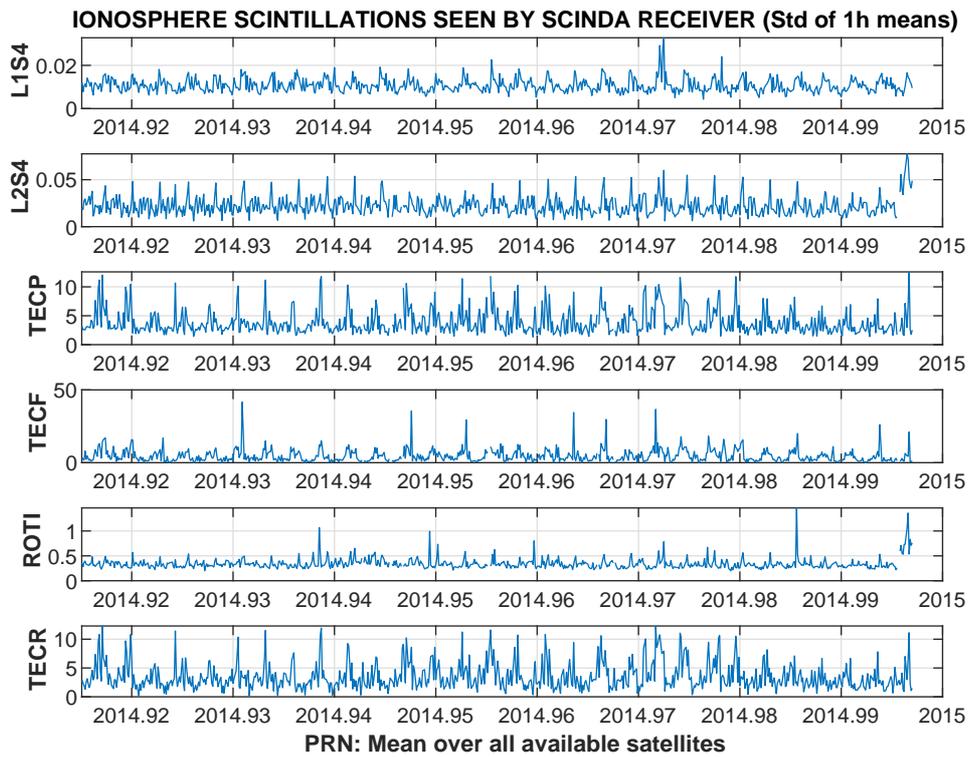}
\vspace{0.5cm}
\caption{Example of standard deviation plots for 1-hour means calculated over an average over all available satellites.}
\label{fig:Mean1h_Std}
\end{figure}

\begin{figure}
\noindent
\centering
\includegraphics[width=0.73\textwidth]{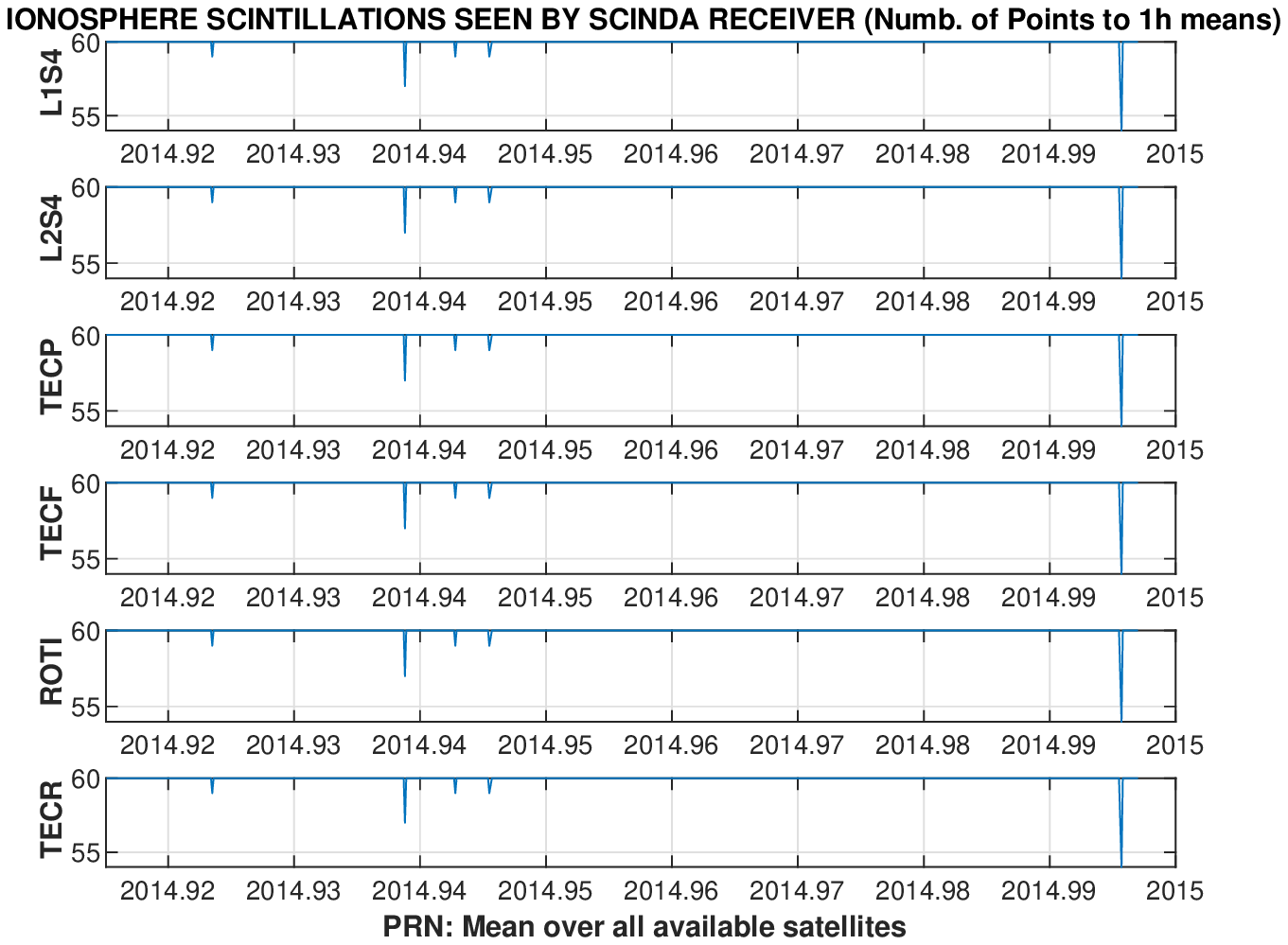}
\vspace{0.5cm}
\caption{Example of number of successful observations plots for 1-hour means calculated over an average over all available satellites.}
\label{fig:Mean1h_Nobs}
\end{figure}

In addition, the 1-hour means of the ``mean'' case and for each receiver-satellite pair are calculated.
The resulting set of 1-hour resolution plots for an average over all available satellites is presented in Fig.~\ref{fig:Mean1h}.
For this resolution the main figure is accompanied by the plots of standard deviation (Fig.~\ref{fig:Mean1h_Std}) and the number of successful observations (Fig.~\ref{fig:Mean1h_Nobs}).

The total time of the data preprocessing, processing and analysis takes, in general, not more than few minutes
in the MATLAB R2018b on the Dell laptop with processor Intel(R) Core(TM) i5-3337U CPU @ 1.80GHz; installed 8,00 GB of RAM; with the 64-bit Operating system (Windows 7 Professional edition).

\section{Impact}
\label{Imp}

The ``SCINDA-Iono'' toolbox can serve as a tool to preprocess and analyze the SCINDA GNSS receiver scintillation data directly and in a fast way.

\begin{itemize}

\item The toolbox simplifies preprocessing and analysis of the ionosphere data that are needed for ionosphere studies.

\item The software can be used for the post-processing of the SCINDA receiver data, to see directly and in real-time the ionosphere conditions shown by these data.

\item The toolbox helps to accelerate and simplify an analysis of ionospheric data.
It can be useful both for scientific tasks and in the applications (in GNSS monitoring and in the analysis needed, for instance, for GNSS Space Weather alert systems).

\item This software has no specific requirements for the user, except basic knowledge of MATLAB tool, and is oriented to the scientific community.

\item The toolbox can be used also in the applied sciences in such areas as GNSS integrity and safety of life services,
as a tool for ionosphere conditions' analysis to be used in GNSS monitoring and for a space weather alert systems.

\end{itemize}

\section{Conclusions}
\label{Concl}

The main goal of the ``SCINDA-Iono'' toolbox is to simplify an analysis of the ionosphere scintillation data provided by the SCINDA GNSS receiver.
Its main functionalities are preprocessing and an analysis of the data on different time-scales and inside the chosen time intervals.
The software can simplify and accelerate the work of ionosphere scientists, as well as can be widely used for the ionosphere data analysis to be used for GNSS space weather alert systems,
to improve the GNSS integrity and safety of life aspects.

For the future work we are planning to extend this toolbox by the analysis of various data formats provided by various models of GNSS receivers.

\section{Conflict of Interest}

We wish to confirm that there are no known conflicts of interest associated with this publication and there has been no significant financial support for this work that could have influenced its outcome.

\section*{Acknowledgements}
\label{Ackn}

The authors thank ARTES IAP DEMOSTRATION PROJECTS and CITEUC founded by FCT, FEDER, COMPETE2020
(UID/MULTI/00611/2019; POCI-01-0145-FEDER-006922).

\newpage

\section*{Current executable software version}
\label{}

\begin{table}[!h]
\begin{tabular}{|l|p{6.5cm}|p{6.5cm}|}
\hline
\textbf{Nr.} & \textbf{(Executable) software metadata description} & \\
\hline
S1 & Current software version & v1.0.1 \\
\hline
S2 & Permanent link to executables  & $https://fr.mathworks.com/$ \\
   & of this version                & $matlabcentral/fileexchange/$ \\
   &                                & $71784-scinda-iono\_toolbox$ \\
\hline
S3 & Legal Software License & MathWorks File Exchange: T.~Barlyaeva personal license \\
\hline
S4 & Computing platforms/Operating Systems & Microsoft Windows \\
\hline
S5 & Installation requirements \& dependencies & MATLAB R2018b \\
\hline
S6 & If available, link to user manual -  & $README\_Barlyaeva\_SCINDA-$ \\
   & if formally published include        & $Iono\_toolbox.txt$ file provided \\
   & a reference to the publication       & with the toolbox at: \\
   & in the reference list                & $https://fr.mathworks.com/$ \\
   &                                      & $matlabcentral/fileexchange/$ \\
   &                                      & $71784-scinda-iono\_toolbox$ \\
\hline
S7 & Support email for questions & TVBarlyaeva@gmail.com \\
\hline
\end{tabular}
\caption{Software metadata} 
\label{}
\end{table}

\end{document}